\documentclass[aps,pra,onecolumn,reprint,showpacs,superscriptaddress]{revtex4-2}
\usepackage{amsmath,amssymb,graphicx,color}
\usepackage{times}
\usepackage{latexsym}
\usepackage{stmaryrd}
\usepackage[table]{xcolor}
\usepackage[colorlinks,allcolors=blue]{hyperref}
\usepackage{amsfonts}
\usepackage{natbib}
\usepackage{appendix}
\usepackage{dsfont}
\usepackage{braket}
\usepackage[english]{babel}

\begin{document}

\title{Passive error correction with a qubit-oscillator system in noisy environment}

\author{Yanzhang Zhu}
\affiliation{Division of Natural and Applied Sciences, Duke Kunshan University, Kunshan, Jiangsu, 215300 China}
\author{Myung-Joong Hwang}%
\email{myungjoong.hwang@duke.edu}
\affiliation{Division of Natural and Applied Sciences, Duke Kunshan University, Kunshan, Jiangsu, 215300 China}
\affiliation{Zu Chongzhi Center for Mathematics and Computational Science, Duke Kunshan University, Kunshan, Jiangsu, 215300 China}

\date{\today}

\begin{abstract}
In this paper, we study an open quantum system consisting of a qubit coupled to a harmonic oscillator subject to two-photon relaxation and demonstrate that such a system can be utilized to construct a cat qubit capable of passive error correction. To this end, we first show that the steady state of the qubit-oscillator system, described by the open quantum Rabi model with two-photon relaxation, undergoes a superradiant phase transition that breaks the strong symmetry of the Lindblad master equation. In the strong symmetry-broken phase, we show that a cat qubit can be stabilized in the steady state by tuning the qubit-oscillator coupling strength and demonstrate that passive error correction can be realized against errors due to fluctuations in the system frequencies. Our study deepens the understanding of dissipative phases in a qubit-oscillator system with strong symmetry and paves the way to utilize them for passive error correction.
\end{abstract}

\maketitle


\section{\label{sec:intro}Introduction}
Quantum information encoded in a quantum system is fragile and susceptible to errors due to inevitable interactions with its environment~\cite{gardiner2004quantum,breuer2002theory,rivas2012open}. This makes error correction one of the most important challenges in building fault-tolerant quantum computers~\cite{nielsen2010quantum,campbell2017roads}. Quantum error correction schemes that rely on measuring error syndromes offer a pathway to achieving fault tolerance, but they require a significant overhead in the number of physical qubits needed to construct a logical qubit, making it challenging to build scalable and practical quantum computers~\cite{campbell2017roads,terhal2015quantum,knill1997theory,li2021statistical}. To circumvent this challenge, passive error correction schemes have been proposed, where quantum errors can be corrected through engineered dissipative processes that interact with the noisy environment~\cite{kerckhoff2010designing,lihm2018implementation,lieu2020symmetry,buvca2019non,kempe2001theory,liu2024dissipative,knill2000theory,gertler2021protecting,herold2017cellular,bombin2015single,bombin2013self}. A bosonic qubit, in which quantum information is encoded in the phase space of a quantum harmonic oscillator, offers a promising qubit architecture for hardware-efficient passive error correction~\cite{mirrahimi2014catqubit,gertler2021protecting,kapit2015passive,kwon2022autonomous}. Recently, it has been recognized that dissipative phase transitions occurring in driven-dissipative bosonic systems can be used as a resource for passive error correction~\cite{lieu2020symmetry,kapit2015passive,liu2024dissipative}. In particular, dissipative phases where the strong symmetry—a symmetry present only in open quantum systems—is spontaneously broken offers a steady-state structure with passive error correction capabilities, as it forms a noiseless subsystem (NS)~\cite{lieu2020symmetry,lidar2014review,viola2001experimental,zanardi2003topological,holbrook2003noiseless,choi2006method,holbrook2005noiseless,kribs2006quantum}.

A quantum harmonic oscillator strongly coupled to qubits undergoes a phase transition, often referred to as a superradiant phase transition (SPT), where macroscopic spontaneous coherence of the bosonic mode emerges~\cite{Emary.2003nzl,Lambert2004,dimer2007proposed,hwang2015quantum}. When the harmonic oscillator is subject to loss, the interplay between the coherent interaction and dissipation stabilizes the macroscopic coherence in the steady-state~\cite{dimer2007proposed,Torre.2013hdtq,hwang2018dissipative,Lyu2024}. SPTs have been observed in various platforms such as Bose-Einstein condensates in optical cavities~\cite{Leonard2017Nature,ETH2021PRX,Farokh2021AdvPhys,li2021first}, superconducting circuits~\cite{zheng2023observation}, and trapped-ions~\cite{cai2021}. So far, SPTs studied in various open qubit-oscillator systems, described by a Lindblad master equation~\cite{lindblad1976generators,gorini1976completely,manzano2020short}, have been mostly limited to those that spontaneously break a \emph{weak} symmetry~\cite{buvca2012note,albert2014symmetries,gu2024spontaneous,dimer2007proposed,hwang2018dissipative}, associated with a single symmetry superoperator that commutes with the Lindblad superoperator. The aim of this paper is twofold: (i) to investigate an SPT occurring in a minimal open qubit-oscillator system that possesses a strong symmetry, where both the Hamiltonian and the jump operators in the Lindblad master equation commute with a parity operator, and (ii) to explore the potential of such a qubit-oscillator system in noisy environment to encode quantum information with passive error correction capabilities.

In this paper, we study the open quantum Rabi model (QRM) with two photon relaxation where an oscillator, subject to two photon loss, is strongly coupled to a single qubit~\cite{malekakhlagh2019quantum}. The even-odd parity of the total number excitation is conserved both by the QRM Hamiltonian and the dissipator for the two-photon relaxation, thereby giving the system a \emph{strong} symmetry~\cite{buvca2012note,albert2014symmetries,gu2024spontaneous,lieu2020symmetry}. After establishing a thermodynamic limit in terms of system frequencies and decay rate, we show that the model exhibits a dissipative SPT that spontaneously break the strong parity symmetry. We show that the Liouvillian gap closes at the phase transition and, as a result, the quadruply degenerate steady-states emerge that are necessary to form the NS. Having established the dissipative phase transition of the model, we show that a cat qubit state can be stabilized by tuning the qubit-oscillator coupling strength. Moreover, quantum errors due to fluctuations in system parameters can be passively corrected as long as the error is confined in the superradiant phase. The QRM Hamiltonian has been realized in advanced quantum technological platforms such as superconducting circuits~\cite{Braumüller.2017,Langford.2017} and trapped-ions~\cite{cai2021}, where the two-photon relaxations can also be readily engineered~\cite{Leghtas2015}; therefore, our study offers an experimentally feasible approach to observe a strong symmetry-breaking SPT and to utilize it to stabilize a bosonic code and to perform a passive error correction.

The paper is organized as follows. In Sec.~\ref{sec:model}, we introduce the model and its strong symmetry. In Sec.~\ref{sec:DPT}, we define a thermodynamic limit where a phase transition may occur and perform a mean-field analysis to determine the phase diagram. We then present numerical results based on a fully quantum treatment to show the Liouvillian gap closing and the extended degeneracy of the steady states. In Section \ref{sec:cat}, we propose stabilizing a cat qubit by tuning the coupling strength and demonstrate passive error correction by examining the fidelity between the target state and the error-corrected state. We conclude with a summary in Section \ref{sec:con}.

\section{\label{sec:model}The Open Quantum Rabi Model with two-photon relaxation}
Let us consider an open quantum system consisting of a qubit coupled to a lossy cavity field, described by the open QRM with two-photon relaxation. The Lindblad master equation for the model reads
\begin{eqnarray}
\label{eq:1}
    \dot{\rho} = \mathcal{L}[\rho] = -i[H_{\textrm{Rabi}},\rho] + \kappa \mathcal{D}[a^2]\rho.
\label{eq:master}
\end{eqnarray}
The coherent dynamics is governed by the QRM Hamiltonian,
\begin{eqnarray}
    H_\textrm{Rabi} = \omega_0 a^\dag a + \frac{\Omega}{2} \sigma_z - \lambda (a + a^\dag) \sigma_x,
\end{eqnarray}
where $\omega_0$ and $\Omega$ are the oscillator and qubit transition frequencies, respectively. $\sigma_{x,z}$ are the Pauli matrices. $\lambda$ is the coupling strength between the two-level system and the oscillator. The two-photon relaxation with a damping rate $\kappa$ is governed by the dissipator in Lindblad form with the jump operator $a^2$,
\begin{eqnarray}
\mathcal{D}[a^2]\rho = 2a^2\rho {a^\dag}^2 - {a^\dag}^2 a^2 \rho - \rho {a^\dag}^2 a^2.
\end{eqnarray}
The model possesses a \emph{strong} parity symmetry since both the Hamiltonian $H_\textrm{Rabi} $ and the jump operator $a^2$ commute with a parity operator,
\begin{eqnarray}
\label{eq:parity}
P=e^{i\pi(a^\dagger a +\frac{1}{2}(\sigma_z+1))}
\end{eqnarray}
, namely,
\begin{eqnarray}
[H_\textrm{Rabi} ,P] = [a^2,P]=0.
\end{eqnarray}
If a single photon loss with a jump operator $a$, described by a dissipator $\mathcal{D}[a]\rho = 2a\rho {a^\dag} - {a^\dag} a \rho - \rho {a^\dag} a$, is introduced to the master equation in Eq.~(\ref{eq:1}), the symmetry of the model reduces to a \emph{weak} parity symmetry. This is because the single photon jump operator $a$ no longer commutes with the parity operator, while the Lindblad master equation is still invariant under a parity transformation $a\rightarrow-a$ and $\sigma_-\rightarrow -\sigma_-$. In the absence of the two-photon loss, Ref.~\cite{hwang2018dissipative} demonstrated that the open QRM with a single photon relaxation undergoes a dissipative phase transition that spontaneously breaks the weak parity symmetry. The goal of the next section is to demonstrate that the two-photon relaxation drives a strong symmetry-breaking dissipative phase transition and investigate its critical properties.

\section{\label{sec:DPT} Dissipative phase transition with broken strong symmetry}
\subsection{Thermodynamic limit}
The QRM Hamiltonian $H_\textrm{Rabi}$ exhibits a second-order ground-state phase transition despite the fact that it consists only of a single qubit and an oscillator~\cite{hwang2015quantum,hwang2016quantum}. This so-called finite-component system phase transition occurs in the thermodynamic limit defined by frequency ratios, namely,
\begin{equation}
\eta\equiv\frac{\Omega}{\omega_0}\rightarrow\infty,\quad \frac{\lambda}{\omega_0}\rightarrow\infty
\end{equation}
while a dimensionless coupling strength remains finite,
\begin{equation}
g\equiv \frac{2\lambda}{\sqrt{\omega_0\Omega}}<\infty.
\end{equation}
The dimensionless coupling strength $g$ is a control parameter that drives a phase transition at a critical point $g_c$. 
Moreover, as the frequency ratio $\eta$ plays the role of the system size, a finite-frequency critical scaling behaviors emerge for $\eta<\infty$~\cite{hwang2015quantum,liu2017}. In the case of the open QRM with a single-photon relaxation, the above definition of thermodynamic limit, in which a dissipative phase transition occurs, remains unchanged~\cite{hwang2018dissipative}. Only the position of the critical point $g_c$ is shifted by the decay rate. In a stark contrast, we find that the thermodynamic limit of the open QRM with two-photon relaxation is realized when an additional infinite frequency ratio and an additional control parameter is introduced. Namely, the thermodynamic limit requires a diverging ratio between the cavity frequency and the decay rate,
\begin{equation}
\zeta\equiv\frac{\omega_0}{\kappa }\rightarrow\infty,
\end{equation}
while keeping a dimensionless decay rate to be finite,
\begin{equation}
h\equiv\frac{\eta}{\zeta}=\frac{\kappa\Omega}{\omega^2}<\infty.
\end{equation}
As we will see below, the finite-$\zeta$ adds non-linear term to the mean-field equation of motion that prevents the non-zero order parameter to be stabilized. Therefore, for the open QRM with two-photon relaxation there are two parameters that play the role of the system size, namely, $\eta$ and $\zeta$ and a dissipative phase transition occurs when both of them are infinite.

\subsection{\label{subsec:semi} Mean-field solutions}
\begin{figure}[t]
\centering
    \includegraphics[width=0.45\linewidth]{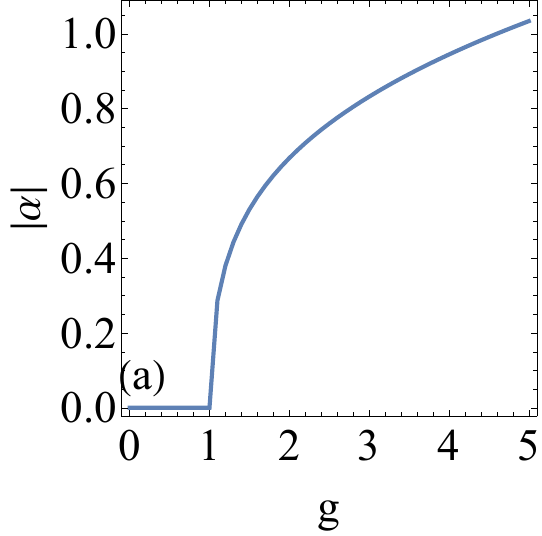}
    \hfill
    \includegraphics[width=0.45\linewidth]{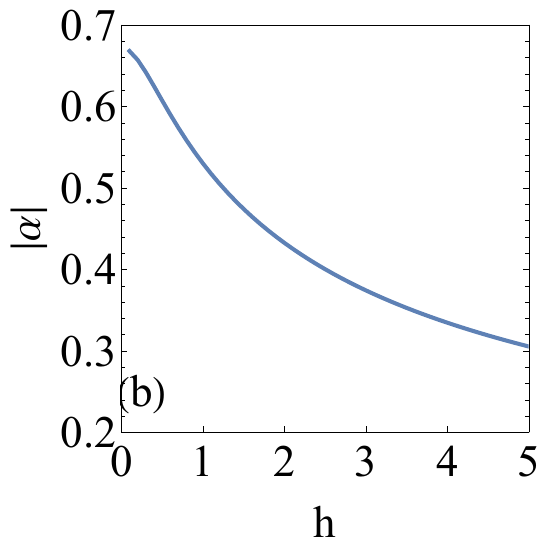}
    \includegraphics[width=0.48\linewidth]{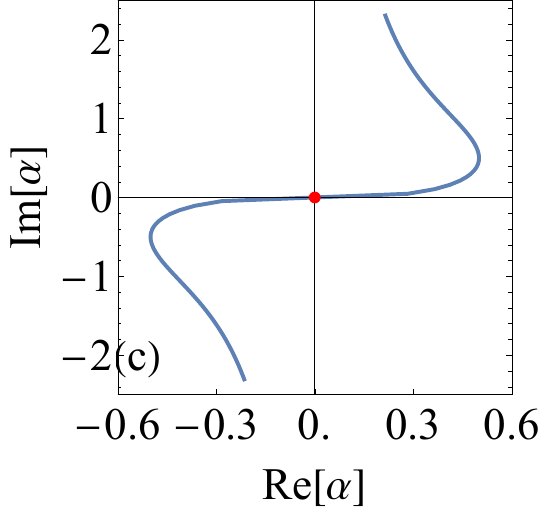}
    \hfill
    \includegraphics[width=0.48\linewidth]{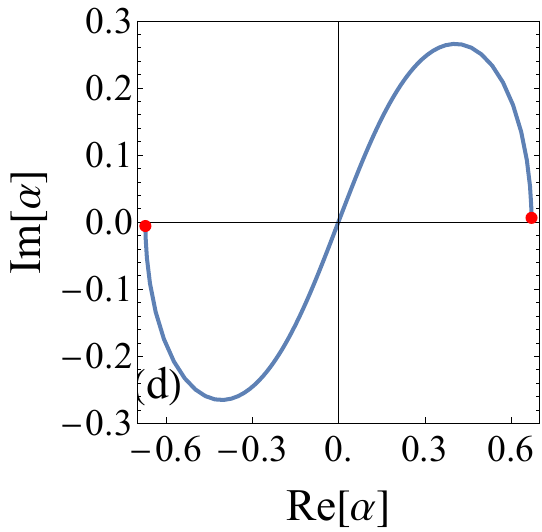}
\caption{\label{fig:order}(a) The magnitude of the order parameter $\lvert \alpha \rvert$ becomes non-zero above a critical coupling strength $g_c=1$. (b) $\lvert \alpha \rvert$ decreases while the renormalized decay rate $h$ increases. (c) Trajectory of $\alpha$ as $g$ increases from $g=1$, which is marked by the red point. (d) Trajectory of $\alpha$ as $h$ increases from $h=0$, marked by the red points. $h=1$ for (a), (c) and $g=1.5$ for (b), (d) are used.}
\end{figure}

We first derive the semiclassical equation of motion of the system,
\begin{eqnarray}
    \braket{\dot{a}} &&= -i\omega_0\braket{a} - i\lambda(\braket{\sigma_+} + \braket{\sigma_-}) - 2\kappa\braket{a^\dag}\braket{a}^2, \nonumber \\
    \braket{\dot{\sigma_+}} &&= i\Omega\braket{\sigma_+} - i\lambda(\braket{a}+\braket{a}^*)\braket{\sigma_z} \nonumber, \\
    \braket{\dot{\sigma_z}} &&= -i2\lambda(\braket{a}+\braket{a}^*)(\braket{\sigma_+} - \braket{\sigma_-}),
\end{eqnarray}
which is obtained from the master equation given in Eq.~(\ref{eq:1}) by applying the mean-field approximation \cite{dutta2019critical,bartolo2016exact,zhang2021driven,hwang2018dissipative}. It is illuminating to cast the above equation into a dimensionless form using a renormalized time $\bar t=\omega_0 t$ and a renormalized order parameter $\alpha\equiv\braket{a}/\sqrt{\eta}$, which leads to
\begin{eqnarray}
    \frac{d}{d\overline{t}}\alpha &&= -i\alpha - i\frac{g}{2} (\braket{\sigma_+} + \braket{\sigma_+}^*) -2 h \lvert \alpha \rvert^2 \alpha, \nonumber \\
    \frac{1}{\eta}\frac{d}{d\overline{t}}\braket{\sigma_+} &&= i\braket{\sigma_+} - i\frac{g}{2}(\alpha+\alpha^*)\braket{\sigma_z}, \nonumber \\
    \frac{1}{\eta}\frac{d}{d\overline{t}}\braket{\sigma_z}&&= -ig(\alpha + \alpha^*)(\braket{\sigma_+} - \braket{\sigma_+}^*).
\label{eq:nonlinear}
\end{eqnarray}
The above equations show that, in the limit of $\eta\rightarrow\infty$, the spin dynamics is completely determined by the order parameter $\alpha$ and therefore can be adiabatically eliminated. In addition, the total spin $\braket{\sigma_x}^2 + \braket{\sigma_y}^2 + \braket{\sigma_z}^2 = 1$ is a conserved quantity due to the absence of the spin damping. The effective mean-field equation of motion for the order parameter after the adiabatic elimination of the spin becomes
\begin{eqnarray}
    \frac{d}{d\overline{t}}x &&= y - 2h(x^2+y^2)x, \nonumber \\
    \frac{d}{d\overline{t}}y &&= -x + \frac{g^2x}{\sqrt{1+4g^2x^2}} - 2h(x^2+y^2)y,
\label{eq:xy}
\end{eqnarray}
where we have introduced $\alpha=x+iy$. In order to determine the steady-state solutions, thereby the phase diagram, we first find the fixed points of the system by setting the left hand sides of both equations  in Eq.~(\ref{eq:xy}) to be zero. A trivial solution $x=y=0$ is always a fixed point. Our numerical analysis shows that there are also non-trivial fixed points with non-zero $x$ and $y$ that always comes in a pair with the same magnitude $|\alpha|=x^2+y^2$. [See Fig.~\ref{fig:order}.]

We then examine the stability of the fixed point solutions using the Jacobian matrix of the dynamical equation, which reads 
\begin{eqnarray}
\textbf{J} = 
\begin{pmatrix}
    -2h(3x^2+y^2) & 1-4h xy \\
    -1+\frac{g^2}{(1+4g^2x^2)^{3/2}}-4h xy & -2h (x^2+3y^2) 
\end{pmatrix}. 
\label{eq:jaco}
\end{eqnarray}
For the trivial fixed point $x=y=0$, the eigenvalue of the Jacobian matrix is given by $\pm \sqrt{g^2-1}$. For $g>1$, one of the eigenvalues become a real and positive value; therefore, it becomes unstable. The stability analysis for $g<1$ requires more careful consideration since the eigenvalues become purely imaginary. In this marginal case of the zero eigenvalues, the linearization method may fail to predict the stability of the fixed points of two-dimensional non-linear dynamical equations \cite{strogatz2018nonlinear}. In the Appendix~\ref{appendix}, we show that the trivial solution is a stable fixed point for $g<1$ using Lyapunov function method. The stability of the non-trivial solutions is analysed by numerically solving the Jacobian matrix, which shows that they are stable only for $g>1$. In Fig.~\ref{fig:order} (a), we plot the magnitude of stable solution $\alpha$ as a function of $g$. Interestingly, we find that the critical point is always $g_c=1$, regardless of the renormalized decay rate $h$ as long as it remains finite. Beyond the critical point, a superradiant phase emerges with non-zero order parameter, indicating the spontaneous breaking of the strong symmetry of the Lindblad master equation.

In Fig.~\ref{fig:order} (b)-(d), we show the dependence of the magnitude of the order parameter $\alpha$ as a function of $h$ for $g>1$. While it remains to be non-zero for any finite values of $h$, the magnitude decreases as the dimensionless decay rate increases. It is interesting to note that in the limit of $h\rightarrow\infty$, the magnitude of $\alpha$ converges to zero, indicating the absence of the superradiant phase and the dissipative phase transition. Since we have already taken the infinite $\eta$ limit for the adiabatic elimination of the spin, the infinite $h$ is realized only when $\zeta$ is finite. In other words, the thermodynamic limit in which the open QRM with two-photon relaxation undergoes a dissipative phase transition requires both $\eta$ and $\zeta$ to be infinite.

\subsection{Liouvillian gap closing and quadruple degeneracy}
A quintessential property of a dissipative phase transition is the closure of the Liouvillian gap, which is defined as the difference between the real part of the eigenvalue with the smallest non-zero real part and the eigenvalue corresponding to the steady state (which has a zero real part) of the Liouvillian superoperator. Let us denote $\Lambda$ as the eigenvalue of the Liouvillian superoperator with the smallest non-zero real part, then the Liouvillian gap is given by $\textrm{Re}[-\Lambda]$. Moreover, a strong-symmetry breaking dissipative phase transition features additional steady-state degeneracy associated with the strong symmetry~\cite{lieu2020symmetry}. In our model, there are two parity superoperators corresponding to the parity operator given in Eq.~(\ref{eq:parity}) that act on either the left and right hand side of an operator. Since both the left and right parity superoperator commute with the Liouvillian superoperator given in Eq.~(\ref{eq:1}), the Liouvillian superoperator can be block-diagonalized into four subspace, $\{ee,oo,eo,oe\}$, depending on the even (e) or odd (o) parity eigenvalues for the left and right parity operators. Each of the $ee$ and $oo$ subspace features a steady-state solution with zero eigenvalue, leading to a doubly degenerate steady-state in the normal phase. In the strong-symmetry broken phases, each one of the subspaces $\{ee,oo,eo,oe\}$ gives rise to steady-state solutions with zero eigenvalues, leading to the quadruple steady-state degeneracy.  

Since the mean-field solution given in the previous subsection only provides information about an average quantity, neglecting the quantum fluctuations, the investigations for the dissipative gap and the degeneracy require going beyond the mean-field analysis. Unfortunately, a numerically exact diagonalization of the Liouvillian superoperator for the Lindblad master equation given in Eq.~(\ref{eq:1}) is forbiddingly difficult due to a large number of basis needed. Our strategy to demonstrate both the closing of the Liouvillian gap and the quadruple degeneracy of the steady-state is to utilize the presence of the two infinite frequency ratios $\eta$ and $\zeta$. Let us consider a limit where $
\eta$ tends to infinity, while $\zeta$ remains to be finite. This limit allows us to exactly integrate out the spin degree of freedom using the Schrieffer-Wolff transformation~\cite{hwang2015quantum}. Namely, we apply a unitary transformation  $U=\exp[g\sqrt{\eta^{-1}}/2(a+a^\dag)(\sigma_+ - \sigma_-)]$ to the master equation in Eq.~(\ref{eq:1}), followed by a projection to the spin down subspace. Note that the spin up subspace also features a stable steady-state, but it is an artifact of assuming the zero spin damping rate. Therefore, we ignore the steady-state in the spin up subspace, which becomes unstable in the presence of realistic spin damping~\cite{Lyu2024}. The effective master equation for the oscillator then becomes
\begin{eqnarray}
    \dot{\rho_a} = &&\mathcal{L}_{\textrm{eff}}[\rho_a]=-i[H_{\textrm{eff}},\rho_a] + \kappa \mathcal{D}[a^2]\rho_a,
\label{eq:effmaster}
\end{eqnarray}
where
\begin{eqnarray}
    H_{\textrm{eff}} = && \omega_0(1-\frac{g^2}{2})a^\dag a - \frac{\omega_0 g^2}{4}(a^2 + {a^\dag}^2),
\end{eqnarray}
and 
\begin{eqnarray}
    \rho_a = \bra{\downarrow}U^\dag \rho U\ket{\downarrow}.
\end{eqnarray}

\begin{figure}[t]
    \centering
    \includegraphics[width=0.47\linewidth]{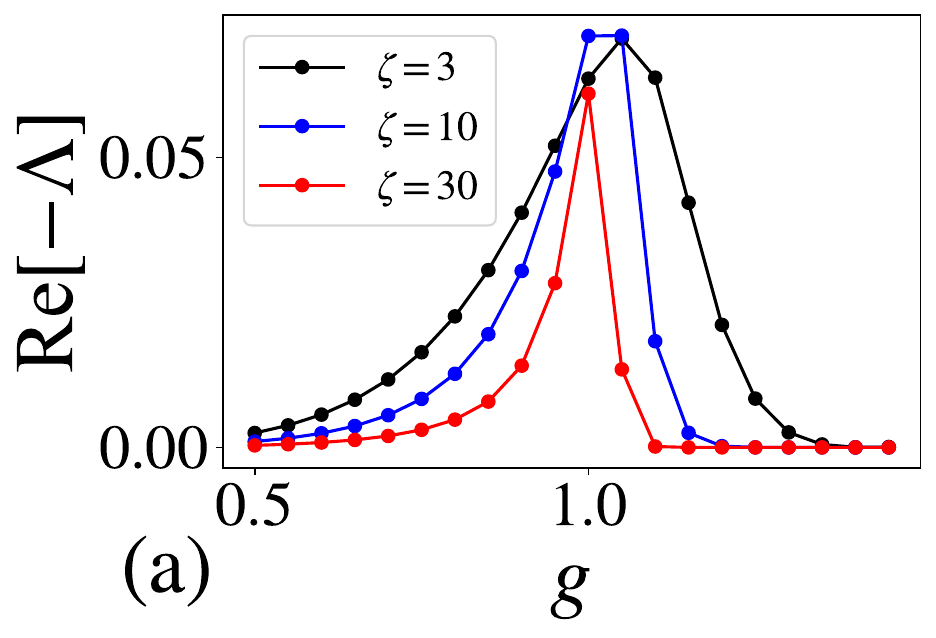}
    \includegraphics[width=0.47\linewidth]{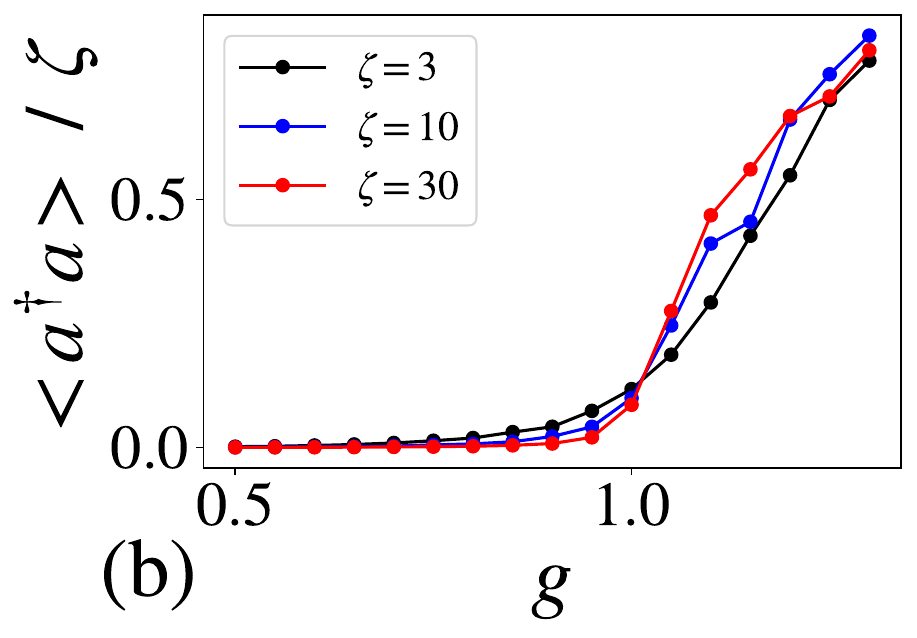}
    \caption{(a) Liouvillian gap, $\textrm{Re}[-\Lambda]$, as a function of g for various values of $\zeta$. The point at which the Liouvillian gap closes occurs for $g>1$ and becomes closer to the critical point $g=1$ as $\zeta = \omega_0/\kappa$ increases. There are two eigenstates corresponding to each point; thus, a 2-to-4 degeneracy transition occurs when the Liouvillian gap closes. (b) The renormalized photon number $\braket{a^\dag a}/\zeta$ becomes nonzero above the critical point $g_c=1$.}
    \label{fig:gap}
\end{figure}

We numerically diagonalize $\mathcal{L}_{\textrm{eff}}$ as a function of $g$ for various values of $h$, which results in two steady-state solutions with zero eigenvalues in $ee$ and $oo$ subspace. In Fig.~\ref{fig:gap} (b), we show that the renormalized photon number expectation value $\braket{a^\dagger a }/\zeta$ of the steady-states sharply becomes non-zero at $g=1$, agreeing with our mean-field prediction. Moreover, the transition becomes progressively sharper as one increases the value of $\zeta$. In Fig.~\ref{fig:gap} (a), we plot the Liouvillian gap, $\textrm{Re}[-\Lambda]$. We note that we obtain two degenerate non-zero eigenvalues that give rise to the same Liouvillian gap. We see that for a large value of $g>1$ the Liouvillian gap becomes exponentially suppressed. As one increases the value of $\zeta$, the gap closure occurs at a value that is progressively closer to the critical point $g_c=1$. Our numerical results demonstrate that the essential features of the dissipative phase transition of the open QRM with two-photon loss emerges even with a moderately large value of $\zeta$.

\section{Passive Error Correction for Cat Qubit}
\label{sec:cat}
The effective master equation for the open QRM with two-photon relaxation, given in Eq.~(\ref{eq:effmaster}), exhibits a special feature at $g=\sqrt{2}$ where the $a^\dag a$ term vanishes. First, coherent states $\ket{\pm\beta}$ with $\beta=\sqrt{\zeta g^2}/2$, defined as $a\ket{\pm\beta}=\pm\beta\ket{\pm\beta}$,  become the steady-states, as well as any arbitrary superposition states between them, forming a decoherence free subspace (DFS)~\cite{mirrahimi2014catqubit,lidar1998decoherence}. s Therefore, the even and odd parity Schrödinger-cat states, 
\begin{eqnarray}
\ket{\beta}_{e,o} = \frac{\ket{\beta} \pm \ket{-\beta}}{2\sqrt{1\pm e^{-2\lvert \beta \rvert^2}}},
\end{eqnarray}
can be stabilized in the steady-state at this point. This so-called cat qubit code has been proven to be a promising candidate for a logical qubit~\cite{mirrahimi2014catqubit}. Second, we note that $g=\sqrt{2}$ is in the strong symmetry-broken phase as demonstrated in the previous section, which forms a noiseless subspace where a passive error correction for cat qubits can be implemented~\cite{lieu2020symmetry}.

Motivated by these observations, we consider a following protocol to demonstrate that a passive error correction for the cat qubit can be realized using our model. We first tune the dimensionless coupling strength at $g=\sqrt{2}$ to stabilize a cat qubit state $\ket{\psi} = c_{e}\ket{\beta}_e + c_{o}\ket{\beta}_o$ in the steady-state. Let us call the Liouvillian superoperator at this operating point as a target Liouvillian $\mathcal{L}_\textrm{target}$ and the prepared state as a target state $\rho_\textrm{target}$. The coefficients $c_e$ and $c_o$ can be determined by a choice of initial state~\cite{mirrahimi2014catqubit}. We then consider an error due to a fluctuation in the dimensionless coupling strength $g$. In the trapped-ion realization of the QRM \cite{pedernales2015quantum,puebla2017,cai2021}, for example, the fluctuation in the frequencies and amplitudes of the Raman sideband lasers gives rise to errors in $\omega_0$, $\Omega$, and $\lambda$. Since $g$ depends on all of the three frequencies, the fluctuation in $g$ could indeed be a dominant source of errors for the open QRM. Therefore, we consider a time-evolution with the Liouvillian with error, $\mathcal{L}_{\textrm{err}}$, where the coupling strength $g$ is drifted away from the operating point for a period of time $\tau$. This introduces an error to the qubit state, leading to $\rho_{\textrm{err}} = \exp[\tau \mathcal{L}_{\textrm{err}}](\rho_\textrm{target})$. The passive error correction is achieved when a time evolution with $\mathcal{L}_\textrm{target}$ restores the target state, thereby autonomously correcting the error. The benchmark for the passive error correction~\cite{lieu2020symmetry} can be achieved by using the fidelity between the error-corrected state, $\rho_\textrm{corr} = \lim\limits_{t \to \infty}\exp[t\mathcal{L}_i](\rho_{\textrm{err}})$ and the target state $\rho_\textrm{target}$, i.e., 
\begin{eqnarray}
F(\rho_\textrm{target}, \rho_\textrm{corr}) = \textrm{Tr}[\sqrt{\sqrt{\rho_\textrm{target}}\rho_\textrm{corr}\sqrt{\rho_\textrm{target}}}]^2.
\end{eqnarray}

\begin{figure}[t]
    \centering
    \includegraphics[width=0.45\linewidth]{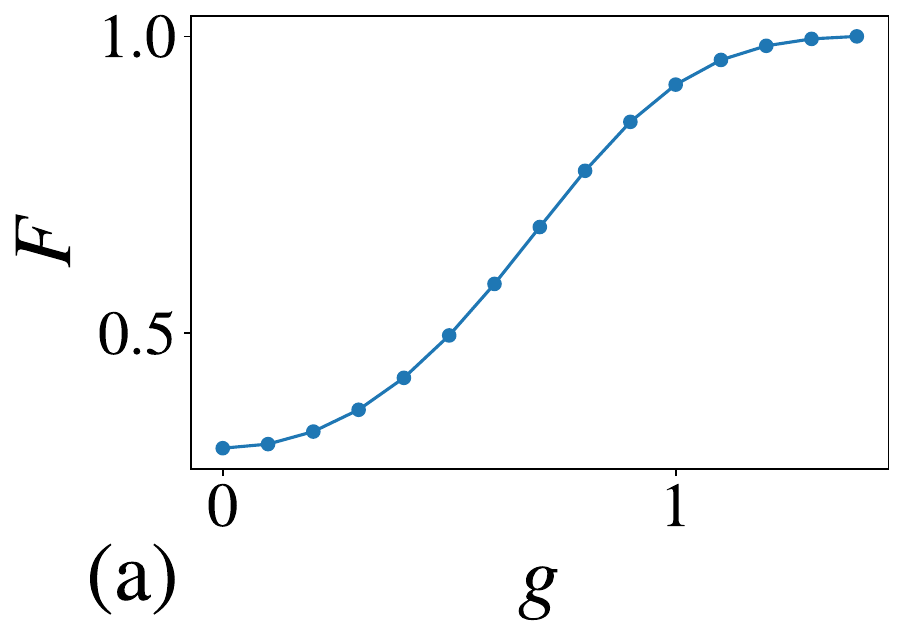}
    \includegraphics[width=0.45\linewidth]{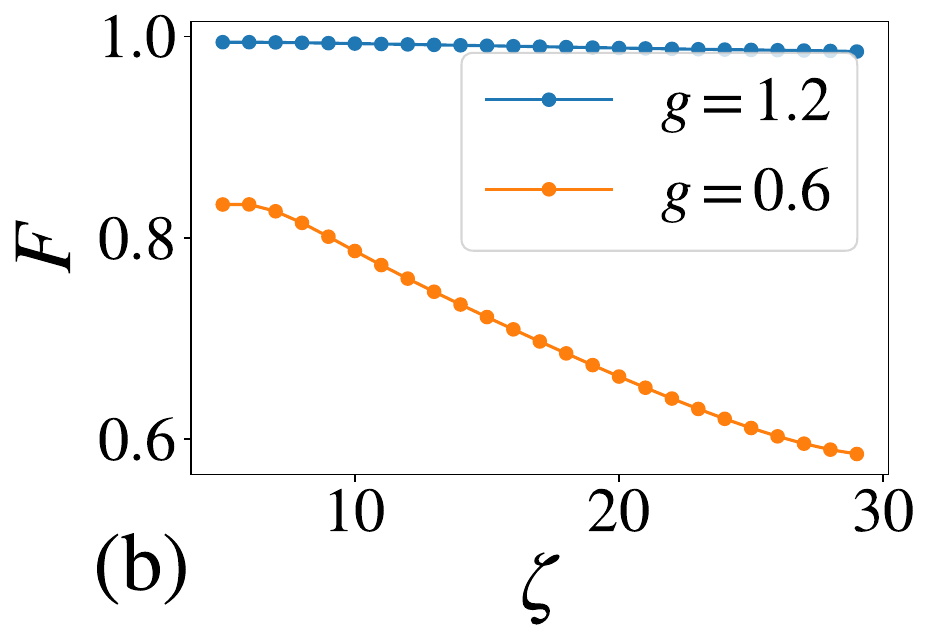}
    \caption{(a) Fidelity increases as $g$ increases, and approaches 1 after the critical point $g_c=1$. (b) As $\zeta \to \infty$, fidelity does not converge to $1$ in the superradiant phase. The quench time is $\tau = 1/\omega_0$ for both (a)(b). The error correction evolution time is $t=1/\omega_0$ for both (a)(b). $\zeta=30$ for (a).}
    \label{fig:fidelity}
\end{figure}

In Fig.~\ref{fig:fidelity}, we present the numerical results for the fidelity $F(\rho_\textrm{target}, \rho_\textrm{corr}) $ obtained for a protocol introduced above. As shown in the Fig.~\ref{fig:fidelity} (a), the fidelity reaches unity only when $g>1$ and decay rapidly as $g$ decreases below the critical point. This demonstrates that the passive error correction for the cat qubit can be indeed achieved in the strong symmetry-broken phase. In Fig.~\ref{fig:fidelity} (b), we show the fidelity as a function of $\zeta$ for both $g<1$ and $g>1$ cases. We note that, for $g<1$, the fidelity after a fixed evolution time $t$ under the target Liouvillian decreases as one increases $\zeta$. While the fidelity for all values of $\zeta$ increases as one increase the evolution time $t$, the fidelity always monotonically decreases as a function of $\zeta$. This result suggests that in the thermodynamic limit of $\zeta\rightarrow\infty$ the fidelity reaches zero for the normal phase ($g<1$) and unity for the broken symmetry phase ($g>1$).

\section{Conclusion}
\label{sec:con}
In this paper, we have studied the open quantum Rabi model with two-photon relaxation as a minimal model that undergoes a SPT, spontaneously breaking a strong parity symmetry. We have demonstrated that the closing Liouvillian gap and the quadruply degenerate steady-states emerge in the superradiant phase. Moreover, our proposal for a passive error correction based on a strong symmetry-breaking SPT paves the way to realize a stabilized bosonic cat code by utilizing a qubit-oscillator coupling, distinct from previous proposals based on parametric two-photon driving~\cite{mirrahimi2014catqubit,lieu2020symmetry} and Kerr nonlinearity~\cite{gravina2023critical}. Given that the mean-field equations of motion for the open Dicke model with two-photon relaxation, which realizes a conventional thermodynamic limit with an infinite number of qubits, can be made identical to the current model after proper renormalization~\cite{hwang2018dissipative,Lyu2024}, we anticipate that the properties and applications of the phase transition we discovered for a single oscillator-qubit system can be generalized to an oscillator coupled to an ensemble of qubits, making our study relevant for a wide range of quantum systems.

\begin{acknowledgments}
This research is supported by the Kunshan Municipal Government research fund.
\end{acknowledgments}

\appendix
\setcounter{figure}{0}
\renewcommand{\thefigure}{A\arabic{figure}}

\section{Stability Analysis with Lyapunov Function}
\label{appendix}

\begin{figure}[b]
    \centering
    \includegraphics[width=0.6\linewidth]{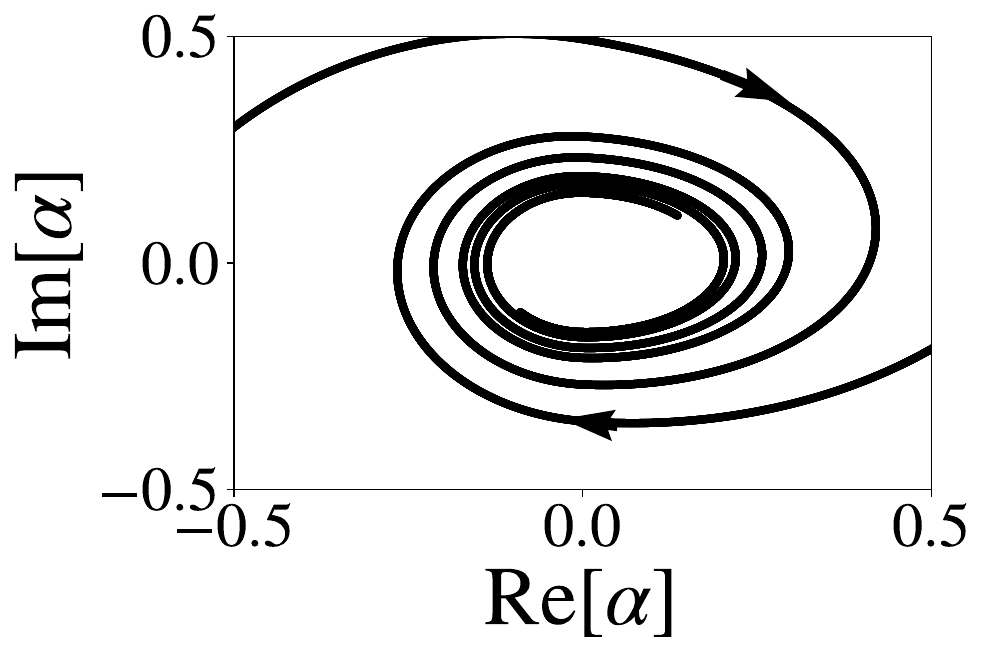}
    \caption{When $g<g_c$, for any arbitrary initial values of $\alpha$, the trajectory of $\alpha$ always goes to the normal phase stable solution $\alpha=0$, following Eq.~(\ref{eq:nonlinear}). Here we take $g=0.6, h=0.25$.}
    \label{fig:simulation}
\end{figure}

In Sec.~\ref{sec:DPT}, we showed that the conventional linearization approach fails to predict the stability of the normal phase, as the real part of eigenvalues of Jacobian matrix becomes zero. In such a marginal case, alternative methods going beyond the linearization is needed to determine the stability of a fixed point~\cite{strogatz2018nonlinear}. In this section, we employ the Lyapunov function method to prove the stability of the normal phase solution. The Lyapunov function is an energy-like function that can be used to prove that any points away from a stable fixed point will decay to it after the time evolution~\cite{vu2005common,strogatz2018nonlinear}. To prove that $(x,y) = (0,0)$ is the only stable fixed point, there are three conditions that a Lyapunov function $V(x,y)$ should satisfy:
\begin{enumerate}
    \item $V(x,y)$ is locally Lipschitz continuous.
    \item $V(x,y)=0$ when $(x,y)=(0,0)$, and $V(x,y)>0$ for all $(x,y)\not=(0,0)$.
    \item $\dot{V}(x,y)<0$ when $(x,y)\not=(0,0)$.
\end{enumerate}

Let us now construct such a Lyapunov function for the normal phase solution of Eq.~(\ref{eq:xy}) as
\begin{eqnarray}
    V(x,y) &&= \frac{1}{2}(x^2+y^2)-\frac{1}{4}\left[\sqrt{1+4g^2x^2}-1\right].
\end{eqnarray}
It is straightforward to check that the condition $1$ and $2$ are satisfied by taking partial derivatives of $V(x,y)$. For condition $3$, we first obtain the time derivative of $V(x,y)$,
\begin{eqnarray}
    \dot{V}(x,y) =&& x\dot{x} + y\dot{y} - \frac{\dot{x}}{4} \frac{d}{dx}F(x) \nonumber \\
    =&&-2h(x^2+y^2)y^2 \nonumber \\
    &&+ 2h(x^2+y^2)(\frac{g^2}{\sqrt{1+4g^2x^2}}-1)x^2,
\end{eqnarray}
where the expressions for $\dot{x},\dot{y}$ from Eq.~(\ref{eq:xy}) are used. Since $g<1$, we find that $\frac{g^2}{\sqrt{1+g^2x^2}}<1$ for $\forall x \not= 0$. When $(x,y)\not=(0,0)$, $x$ and $y$ cannot be $0$ simultaneously, thus $\dot{V}(x,y)<0$. Condition 3 is therefore satisfied. Since there exists a function $V(x,y)$ that satisfies all three conditions, $(x,y)=(0,0)$ is the only stable point for $g<1$. 

Finally, we confirm that the normal phase is indeed a stable solution by numerically solving the coupled differential equations in Eq.~(\ref{eq:xy}). We present a phase portrait in Fig.~\ref{fig:simulation}, which shows a spiral trajectory toward the fixed point at origin. Note that the decay time toward the fixed point is extremely slow, which is commonly observed for the stable fixed point with a marginal eigenvalue~\cite{strogatz2018nonlinear}.

\bibliography{paper}

\end{document}